%% file: Gravitino_Trilinear.tex
\RequirePackage[displaymath]{lineno}
\documentclass[a4paper,11pt]{article}
\pdfoutput=1 

\usepackage{jcappub}
\usepackage[T1]{fontenc} 

\usepackage{cancel}

\usepackage{aas_macros}
\usepackage{tablefootnote}
\usepackage[displaymath]{lineno}
\usepackage{subfigure}
\usepackage{multirow}
\usepackage{amsmath}
\usepackage{cleveref}

\usepackage{enumitem}

\usepackage{graphicx}
\usepackage[colorinlistoftodos]{todonotes}

\usepackage{soul}

\DeclareUnicodeCharacter{2212}{-}

\notoc

\begin{document}

\title{Probing the anomalous positron fraction origin with fully leptonic decaying gravitino dark matter candidates}

\input{authors.tex}


\abstract{The flux of electron and positron cosmic rays measured by the space-based experiments PAMELA, AMS-02, CALET, DAMPE and Fermi-LAT shows an unexpected behaviour at high energies, in comparison with expectations from standard astrophysical sources. In particular, AMS-02 observations provide compelling evidence for a new source of positrons and electrons whose origin is still unknown. Plausible scenarios include either the contribution of dark matter or unresolved astrophysical sources, such as nearby pulsars. It has been shown that explanations based mostly on dark matter, tend to overproduce $\gamma$-rays, entering in conflict with measurements of the extra-galactic $\gamma$-ray background (EGB). Although this situation seems to be quite generic, it ultimately depends on the properties of the dark matter candidate. In this work we revisit a model in which the gravitino is a dark matter candidate decaying to SM particles through R-parity violating couplings. We show that decay channels allowed by trilinear couplings produce more electrons and positrons and fewer photons in comparison to channels allowed by bilinear couplings. Indeed, when considering AMS-02 data alone we find that the trilinear model is compatible with EGB constraints. However, this result may change when DAMPE or CALET measurements of the sum of electrons and positrons fluxes are considered instead of AMS-02 data. Therefore, in order to evaluate this model further, discrepancies between the data collected by these experiments need to be clarified.}


\maketitle

\newpage
\section{Introduction}

The energy spectrum of electrons and positrons measured by experiments such as PAMELA~\cite{Adriani:2008zr}, AMS-02~\cite{Accardo:2014lma,Aguilar:2014mma,Aguilar:2014fea}, CALET~\cite{Adriani:2018ktz}, DAMPE~\cite{Ambrosi:2017wek} and Fermi-LAT~\cite{Ackermann:2014usa} show noticeable discrepancies when compared to predictions based on standard astrophysical sources, such as cosmic-ray interactions or emissions from pulsars. These experimental signals suggest that extra sources of (primary) positrons are required in order to make sense of the data. 

It is well known that an extra source  of cosmic rays can be obtained from Dark Matter (DM) annihilations or decays~\cite{Hooper:2018kfv}. For instance, it has been shown that the anomalous measurement of positrons, initially reported by PAMELA~\cite{Adriani:2008zr} and later confirmed by AMS-02~\cite{Accardo:2014lma}, can be well explained by considering a distribution of DM particles that can annihilate or decay to charged leptons both as primary or secondary particles~\cite{Grefe:2008zz,2012PhRvD..86h3506C,Ando:2015qda,Laletin:2016egv,Liu:2016ngs,Belotsky:2016tja}. This generic picture, however, when looking at the produced gamma ray spectrum, becomes in conflict with the measurement of the extra galactic gamma-ray background (EGB) derived from Fermi-LAT~\cite{Ackermann:2014usa}. As an example about this issue, we refer to our previous work~\cite{Carquin:2015uma}, where we test a DM scenario that contains a gravitino as the lightest super-symmetric particle (LSP) and bilinear R-parity violation (BRpV) couplings~\cite{Grefe:2011dp}. We found that this model is able to describe the anomalous positron fraction measured by AMS-02, but at the cost of producing a flux of $\gamma$-rays overshooting the EGB limits. 

In this work, we continue with the study of gravitino models, basically searching for scenarios where the production of $\gamma$-rays could be reduced to stay away from the EGB constraint. In particular, we suggest that this can be achieved by moving from BRpV to trilinear R-parity violating (TRpV) couplings, since the tree-level gravitino decay modes allowed in each scenario are completely different. For instance, in TRpV the gravitino is allowed to decay to two charged leptons plus a neutrino, such as $\psi \rightarrow l^{\mp}l^{\pm}\nu$ with $l=e,\mu,\tau$, while in BRpV the gravitino decays involve two-body decays including leptons and higgs/gauge bosons, such as $\psi \rightarrow l^{\mp} W^{\pm}$ and $\psi \rightarrow \nu H/Z$. In practice, we have noticed that preferred three-body decays, which are useful to fit the positron data, produce more charged leptons than  two-body decays for the same lifetime, which is helpful to reduce the associated emission of gamma-rays. Although this approach seems to work for the analysis of AMS-02 and the EGB, we notice that the data reported from CALET and DAMPE may have a relevant impact on this conclusion.

The paper is organized as follows. In~\cref{sec:gdecay} we describe the decaying gravitino DM model with R-Parity violating couplings. In~\cref{sec:comparison} we compare trilinear and bilinear decay channels in order to show the features of TRpV that allow a better comparison to cosmic ray data. In~\cref{sec:observables} we define the relevant contributions to the electron, positron and photon flux. In~\cref{sec:stats} we discuss the data considered in the current analysis and the statistical tests of our model. In~\cref{sec:results} we comment on these findings. Finally, in~\cref{sec:conclusions} we summarize our conclusions and perspectives.  
 
\section{Dark Matter model}
\label{sec:gdecay}

We consider a super-symmetric (SUSY) extension of the standard model (SM) with a low energy spectrum characterized by a gravitino as the lightest SUSY particle (LSP), that can decay to final state leptons through R-Parity violating couplings~\cite{Grefe:2011dp, Moreau:2001sr}. In this scenario, the decay of the gravitino LSP can be achieved in two steps. 

First, we consider R-Parity conserved interactions between the gravitino, one SM fermion and the corresponding scalar super-partner, which in principle do not allow the direct decay of the gravitino. If we represent the SM fermions by $\psi$, the scalar super-partners as $\phi$ and the gravitino field as $\Psi_\mu$, we can write the Lagrangian associated to this interaction as:

\begin{equation}
  \mathcal{L} = −\frac{1}{\sqrt{2}M_{*}}\bar{\psi}_{L}\gamma^{\mu}\gamma^{\nu}\partial_{\nu}\phi\Psi_{\mu R}
\end{equation}

\noindent where the $L/R$ indices standing for left/right chirality, $\gamma^\mu$ being the Dirac matrices and $M_* = (8\pi G N)^{-1/2} = 2.4\times 10^{18}$ GeV the reduced Planck mass. In this notation the mass of the gravitino is given by $m_G = F/\sqrt{3}M_*$, with $F$ the scale of spontaneous-SUSY breaking. 

Second, we must consider R-Parity violating interactions, which in practice allow the decay of one single superpartner into final state channels that only contain SM particles. These interactions are modelled by the following superpotential,

\begin{equation}
 W_{\cancel{R}_p} =  \frac{1}{2}\lambda_{ijk}L_i L_j E_k^c + \lambda'_{ijk}L_i Q_j D_k^c + \lambda''_{ijk}U_i D_j D_k^c + \epsilon_iHL_i,
\label{eq:superpotential}
\end{equation}

\noindent which is written in terms of superfields and scalar couplings. This superpotential includes left-handed lepton and quarks superfields ($L$) and ($Q$), the up-type Higgs superfield ($H$), right-handed lepton and quark superfields ($E^c$) and ($U^c$, $D^c$). The indices $i,j,k$ run over flavor generations, $\lambda_{ijk}$, $\lambda'_{ijk}$, $\lambda''_{ijk}$ are dimensionless coupling constants and $\epsilon_i$ are dimension one parameters. 

The bilinear term of Eq.~\ref{eq:superpotential}, controlled by the coupling $\epsilon_i$, determines the strength of two body gravitino decays, such as $\psi\rightarrow H\nu, W^{\pm} l^{\mp}, Z\nu$. In a previous work~\cite{Carquin:2015uma} we have considered these decays in order to adjust the AMS-02 positron anomaly, but we have also found that the model is in tension with EGB limits from Fermi-LAT. Here we consider these results in order to guide the search of viable scenarios. In particular, we focus on the first type of interactions, which are controlled by the couplings $\lambda_{ijk}$, since these terms allow the three-body decay of gravitinos to pairs of opposite charged leptons plus one neutrino, such as $\psi  \rightarrow l^{-}_il^{+}_j\nu_k$. We consider only these terms because the decay channels contain charged leptons, which are useful to adjust the positron anomaly, but also because we can get lower amounts of $\gamma$ radiations in comparison to the cases where we have quarks as final states, which appear when $\lambda_{ijk}'$ and $\lambda_{ijk}''$ are switched on. In ~\cite{Moreau:2001sr}, we can find the analytical expressions for the decay rates for every combination of leptons produced by $\lambda_{ijk}$ terms. In Appendix~A, we also discuss the possibility to use only these terms of the superpotential to explain at least one of the neutrino masses, such as in reference~\cite{Chun:2004mu}. We notice that the heavy mass of the scalars, which is required to make the lifetime of the gravitino sufficiently small, also allow neutrino mass contributions considering reasonable values of trilinear couplings. 

\section{Comparison between BRpV and TRpV scenarios}
\label{sec:comparison}

When we consider a SUSY model that includes BRpV terms~\cite{Grefe:2011dp}, it is possible to have tree-level interactions between a gravitino and two SM particles. Thus, for a gravitino mass above the TeV scale, which is necessary to adjust the positron anomaly~\cite{Aguilar:2014mma}, gravitinos can decay through on-shell two-body final state channels, that can contain gauge bosons and leptons ($\psi\rightarrow Z \nu$ and $\psi\rightarrow W^{\pm} l^{\mp}$) or a Higgs particle plus neutrinos ($\psi\rightarrow H\nu$). In particular, in our previous work~\cite{Carquin:2015uma}, we have shown that the preferred gravitino lifetime in this case is $1.0\,(1.3)\times10^{26}\,\text{s}$ for $m_G = 1\,(2) \,\text{TeV}$. The preferred channel is mostly given by $\psi\rightarrow W^{\pm} \tau^{\mp}$, with 90\% (80\%) of the total branching fraction.

Instead, when we consider TRpV interactions~\cite{Grefe:2011dp}, it is possible to have tree-level interactions between the  gravitino, one SUSY scalar particle and the corresponding SM fermion, which can be obtained from the first term of Eq.~\ref{eq:superpotential}. Thus, when scalars are considered to be super heavy, such as in Split-SUSY models~\cite{Giudice:2004tc} or in order to enhance the gravitino lifetime (Appendix~\ref{sec:lifetime-neutrinomasses}), gravitinos only can decay through off-shell three-body final state channels, which in general include two charged leptons plus a neutrino ($\psi\rightarrow l^{\mp} l^{\pm} \nu$). Notice that the branching fractions of conjugated channels must be equal, such that $Br(\psi \rightarrow l'^{\mp}l^{\pm} \nu) = Br(\psi \rightarrow l'^{\pm}l^{\mp} \nu)$. 

\begin{figure}[htb]
\begin{center}
\includegraphics[height=7cm,width=7cm,angle=0]{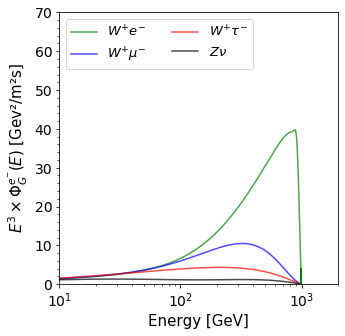}
\includegraphics[height=7cm,width=7cm,angle=0]{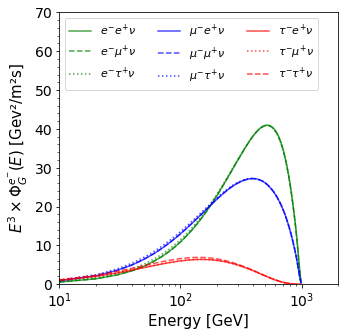}
\caption{Electron spectrum of bilinear (left panel) and trilinear (right panel) decay channels. In both cases we consider those channels that are able to produce prompt electrons. The gravitino lifetime is the same and equal to $\tau_G = 10^{26}\,s$ in both panels.}
\label{fig:electron-spectrum}
\end{center}
\end{figure}

Now let us discuss some features that arise when we consider the predicted spectrum of charged leptons in TRpV models. For simplicity, we focus on the measurement of electrons, but the positron case is analogous and symmetric. In this case we can consider that neutrino flavors are indistinguishable, thus in principle we only have to deal with nine channels, which we can list from 1 to 9 as $\psi\rightarrow l_i^{-}l_j^{+}\nu$ with $i,j=e,\mu,\tau$. We must notice that the electron spectrum of each element of the group $l'^{-}l^{+}\nu$ with $l'$ fixed, generates a very similar spectra, which is shown in Fig.~(\ref{fig:electron-spectrum}) right panel. This feature allows us to reduce the number of effective independent channels to adjust the electron-positron data from nine to three. In section~\ref{sec:stats} we use this feature to show that the preferred gravitino lifetime in this case is $4.6\,(3.6)\times10^{26}\,\text{s}$ for $m_G = 1.3\,(2.3) \,\text{TeV}$. The preferred channel is mostly given by $\psi\rightarrow \tau^{-} l_i^{+}\nu$ with $i=e,\mu,\tau$, with 54\% (58\%) of the total branching fraction.

From these results we can see that in TRpV we requires a gravitino lifetime ($3.6\times10^{26}\,\text{s}$) which is slightly larger than the one obtained in BRpV ($1.3\times10^{26}\,\text{s}$). This is because the preferred channel in TRpV ($\tau^{-} l^{+}\nu$) generates relatively more electrons than the preferred channel in BRpV ($W^{+} \tau^{-}$), which can be checked by comparing the red curves in both plots of Fig.~(\ref{fig:electron-spectrum}). This effect is probably due to phase space differences between three-body and two-body decays, which may be interesting to understand further. However, in this work we consider that the visual inspection of the spectra behavior is enough to motivate a further study of the TRpV scenario.

\begin{figure}[htb]
\begin{center}
\includegraphics[height=7cm,width=7cm,angle=0]{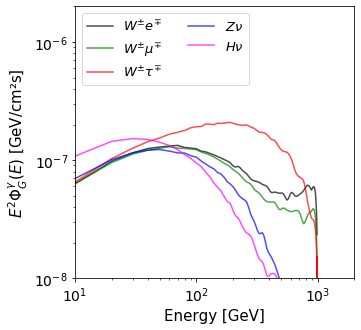}
\includegraphics[height=7cm,width=7cm,angle=0]{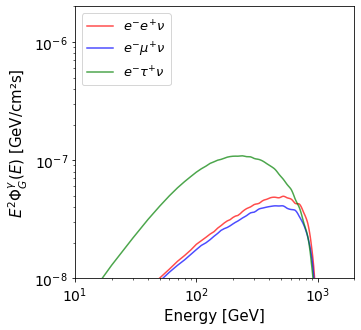} \\
\includegraphics[height=7cm,width=7cm,angle=0]{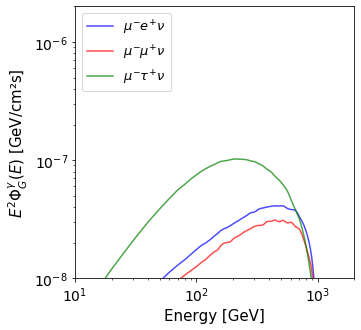}
\includegraphics[height=7cm,width=7cm,angle=0]{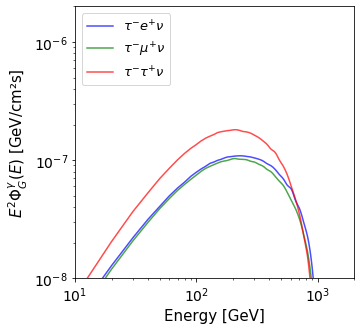}
\caption{Photon spectrum for bilinear and trilinear models. The bilinear case is shown in the top left panel. The  trilinear cases are given by the top right panel, which includes the channels 1, 2 and 3, the bottom left panel, which includes the channels 4, 5 and 6, and the bottom right panel, which includes the channels 7, 8 and 9 (see definitions of these channels in the main text). The gravitino lifetime is the same and equal to $\tau_G = 10^{26}\,s$ in both panels.}
\label{fig:photon-spectrum}
\end{center}
\end{figure}

\begin{figure}[htb]
\begin{center}
\includegraphics[height=7cm,width=7cm,angle=0]{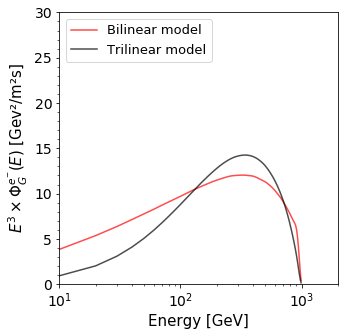}
\includegraphics[height=7cm,width=7.5cm,angle=0]{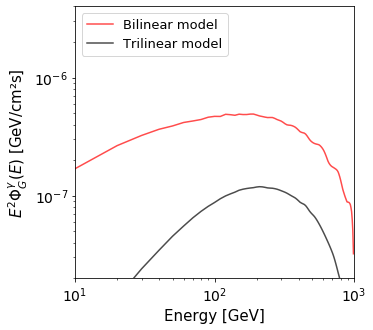}
\caption{Expected electron (left panel) and photon spectrum (right panel) for bilinear and trilinear models. In bilinear scenario we use $m_G = 2$ TeV, $\tau_G=1.3\times10^{26}\, s$ and $BR(\psi\rightarrow W^{+}\tau^{-}) = 80\%$, which correspond to one of the preferred scenarios obtained in Ref.~\cite{Carquin:2015uma}. In trilinear case we use $m_G = 2$ TeV, $\tau_G=3.6\times10^{26}\, s$ and $BR(\psi\rightarrow \tau^{-}l^{+}\nu) = 58\%$, which is one of the best fit scenarios found in this paper.}
\label{fig:best-fit-spectrum}
\end{center}
\end{figure}

Furthermore, we also can check that, for the same gravitino lifetime, the gamma-ray spectra produced by BRpV channel ($W^{+}\tau^{-}$) generates a slightly higher photon flux than the preferred channel of TRpV ($\tau^{-} l^{+}\nu$), which can be checked by comparing the corresponding lines of Fig.~(\ref{fig:photon-spectrum}). Basically, we may conclude that for the given gravitino lifetime, the TRpV scenario generates an enhancement of charged leptons in comparison to BRpV and at the same time the flux of photons is slightly smaller. Mixing these two effects we can see that the TRpV model is able to generate a sizable amount of electrons at a lower photon cost in comparison to BRpV, which can be seen in Fig.~(\ref{fig:best-fit-spectrum}). In the following sections we study the TRpV model considering the data from AMS-02~\cite{Accardo:2014lma,Aguilar:2014mma,Aguilar:2014fea}, CALET~\cite{Adriani:2018ktz}, DAMPE~\cite{Ambrosi:2017wek} and Fermi-LAT~\cite{Ackermann:2014usa} in order to check this scenario with current data and a proper statistical analysis. 

\section{Standard contributions to electron, positron and $\gamma$-rays}
\label{sec:observables}

The standard contribution to the flux of electrons and positrons detected at Earth arise from local galactic sources, such as pulsars or spalls from primary cosmic rays interacting with the interstellar medium. The majority of this flux can be accounted by a diffuse power law spectra~\cite{Aguilar:2013qda}, which is expected from standard astrophysical models. Therefore, we model this contribution with the following expressions:

\begin{eqnarray}
\Phi_B^e(E) &=& C_e (E/1\,\text{GeV})^{-\gamma_e} \hspace{3mm}\text{for electrons and}\\
\newline
\Phi_B^p(E) &=& C_p (E/1\,\text{GeV})^{-\gamma_p} \hspace{3mm}\text{for positrons}
\end{eqnarray}

\noindent where the values of $C_e, C_p, \gamma_e$ and $\gamma_p$ are obtained from the fit to the data. The astrophysical background term in the electron flux attempt to model primary and secondary production, while in the positron flux equation represents only secondary production. However, in order to accommodate the rise of the positron flux above $\approx 200$ GeV, reported by AMS-02~\cite{Accardo:2014lma}, we definitely need something else. In order to cover this gap we include a new source term injecting electrons and positrons at high energies, that arises from the decay of gravitinos into SM particles. The details about this contribution are given in the following section.

The charged SM particles produced from gravitino decays must produce $\gamma$-rays, that would reach specialized detectors, such as Fermi-LAT. Indeed, it has been noticed that the DM emission of $\gamma$-rays at high galactic latitude ($|b| > 20^\circ$) would be indistinguishable from the overall isotropic gamma-ray background (IGRB)~\cite{Hooper:2018kfv}, which is a tenuous diffuse component. Since resolved sources are not part of the IGRB, this observable depends on the point source threshold of the specific experiment. Thus, it is preferable to consider the measurement of the total extra-galactic $\gamma$-ray background (EGB), which represents a superposition of all $\gamma$-ray sources, both resolved and unresolved, from the
edge of the Milky Way to the edge of the observable Universe~\cite{Ackermann:2014usa}. In other words, the EGB is the sum of the IGRB plus resolved sources. It turns out that the majority of the EGB can be associated to the emission of standard sources~\cite{Hooper:2018kfv}. Therefore, we use the EGB to put limits on DM contributions. The standard contributions to the EGB are taken from~\cite{Carquin:2015uma} and references therein.

\section{Gravitino contributions to electron, positron and $\gamma$-rays in TRpV}

On top of the standard astrophysical background we consider the contribution of gravitino decays. Since different neutrinos are indistinguishable, we must deal with 9 independent channels, i.e. $l^{-}_i l^{+}_j\nu$ for $i,j=e,\mu,\tau$. Therefore the relation between branching fractions and individual trilinear couplings is not necessarily direct. We define the branching fractions from $Br_1$ to $Br_9$, such that $Br_1 = Br(\psi \rightarrow e^{-}e^{+}\nu$), $Br_2 = Br(\psi \rightarrow e^{-}\mu^{+}\nu$), etc. Considering these channels we can  model the amount of electrons, positrons or $\gamma$-rays produced on gravitino decays as:

\begin{equation}
\Phi_{G}^{\eta}(E) = \frac{1}{m_G \tau_G} \sum_{j=1}^9 {Br_j \frac{dN_j^{\eta}}{dE}} D^{\eta}_{\text{factor}},
\label{dm-flux}
\end{equation}

\noindent where $m_G$ and $\tau_G$ are the mass and lifetime of the gravitino respectively. The term $\eta = e, p,\gamma$ when we consider electron, positron or $\gamma$-ray flux, correspondingly. The $D^{\eta}_{\text{factor}}$ is proportional to the density of DM in the case of $\eta=\gamma$, in the other cases is a more complex term that depends on the DM density and the propagation of charged particles in the Galaxy. The term $dN_j^{\eta}/dE$ is the amount of electrons, positrons, or $\gamma$-rays produced per gravitino decay and energy of the corresponding particle, propagated to the Earth position.

\subsection*{Effective channels for data analysis: electron-positron}

For the computation of the electron-positron spectrum at Earth's position, we consider an approach similar to that used in our previous work~\cite{Carquin:2015uma}, therefore we suggest to follow that work and references therein, to check the details on the total flux computations, including propagation effects. For simplicity, we restrict the current analysis to the MED propagation working point only.

Now, let's focus on the spectrum of charged leptons. In principle, we can get each branching fraction as a function of the free parameters of the fundamental model, such as the trilinear couplings $\lambda_{ijk}$ and the mass of super-symmetric scalars. However, for our purposes it is sufficient to consider directly the branching fractions as the effective free parameters for the fit of the electron and positron spectrum. Besides, by considering that some of the channels generate a very similar spectrum (see Fig.~\ref{fig:electron-spectrum} right panel), we can reduce the model degrees of freedom further, by grouping decay channels as follows:

\begin{equation}
\Phi_{G}^{e}(E) \propto \frac{1}{m_{G}\tau_{G}}\biggl[\alpha_{1}\frac{dN^e_{1}}{dE}+\alpha_{2}\frac{dN^e_{2}}{dE}+\alpha_{3}\frac{dN^e_{3}}{dE}\biggr],
\end{equation}

\noindent where $\alpha_{1}=Br_{1}+Br_{2}+Br_{3}$, $\alpha_{2}=Br_{4}+Br_{5}+Br_{6}$
and $\alpha_{3}=Br_{7}+Br_{8}+Br_{9}$ with $\alpha_{1}+\alpha_{2}+\alpha_{3}=1$. Thus, we only need to find two independent effective branching fractions
in order to adjust the electron spectrum. Analogously, for the positron spectrum we
have that

\begin{equation*}
\Phi_{G}^{p}(E)  \propto  \frac{1}{m_{G}\tau_{G}}\biggl[\beta_1\frac{dN^{p}_{1}}{dE}+
  \beta_2\frac{dN^{p}_3}{dE}+
  \beta_3\frac{dN^{p}_3}{dE}\biggr],
\end{equation*}

\noindent where $\beta_1 =Br_{1}+Br_{4}+Br_{7}$, $\beta_2 = Br_{2}+Br_{5}+Br_{8}$ and $\beta_3 = Br_{3}+Br_{6}+Br_{9}$. In principle, it seems that we need two extra parameters to fit the positron spectrum, however it is possible to show that electron and positron spectra are indeed equivalent. In order to find this explicitly, we use the equivalence between branching fractions of conjugated decay channels, such that $Br_2 = Br_4$, $Br_3 = Br_7$ and $Br_6 = Br_8$. Thus, we can rewrite the positron spectrum as

\begin{equation*}
    \Phi_{G}^{p}(E) \propto  \frac{1}{m_{G}\tau_{G}}\biggl[\alpha_{1}\frac{dN^p_{3}}{dE}+\alpha_{2}\frac{dN^p_{5}}{dE}+\alpha_{3}\frac{dN^p_{1}}{dE}\biggr]
\end{equation*}

Finally, we can use that the electron spectrum from a given channel must be equal to the positron spectrum of the conjugated one to find that

\begin{eqnarray}
\Phi_{G}^{p}(E) & \propto & \frac{1}{m_{G}\tau_{G}}\biggl[\alpha_{1}\frac{dN^e_{1}}{dE}+\alpha_{2}\frac{dN^e_{2}}{dE}+\alpha_{3}\frac{dN^e_{3}}{dE}\biggr] \\
\Phi_{G}^{p}(E) &=& \Phi_{G}^{e}(E)
\label{ele-pos-spec}
\end{eqnarray}

Therefore, in order to fit the electron-positron data we only need to consider two independent effective branching fractions. Notice that the equivalence between electron and positron spectra is expected from general arguments related to charge conjugation symmetry, here we just wanted to show and verify this issue explicitly in terms of our effective parameters.

\subsection*{Effective channels for data analysis: $\gamma$-rays}

Analogously to the electron-positron flux, we may discuss the total contribution of gravitino decays to the EGB measured at Earth's position by considering the following expression

\begin{equation}
\Phi_{G}^{\gamma}(E) \propto\frac{1}{m_{G}\tau_{G}}\sum_{i=1}^{9}Br_{i}\frac{dN^{\gamma}_{i}}{dE}
\end{equation}

\noindent where $Br_i$ are the same branching fractions that appear in front of the electron-positron spectra. Thus, we can use some results of the previous section in order to manipulate these branching fractions and optimize the contribution of gravitino decays to the EGB. 

Indeed, we have seen that the spectrum of charged leptons effectively depends on the parameters $\alpha_1$, $\alpha_2$ and $\alpha_3$, which are given by the sum of individual branching fractions. Therefore, after the fit of the electron-positron data, the individual branching fractions are still undetermined. Basically, we are free to choose $BR_{1}$ to $BR_{9}$ subject to the conditions, $\alpha_{1}=Br_{1}+Br_{2}+Br_{3}$, $\alpha_{2}=Br_{4}+Br_{5}+Br_{6}$ and $\alpha_{3}=Br_{7}+Br_{8}+Br_{9}$. Besides, we also must satisfy that $Br_2 = Br_4$, $Br_3 = Br_7$ and $Br_6 = Br_8$. Considering these conditions and the particular shapes of the $\gamma$-ray spectrum in our model (see Fig.~\ref{fig:photon-spectrum}), it is possible to show that the combination that minimizes the total $\gamma$-ray spectrum is given by

\begin{eqnarray}
Br_{1} & = & \alpha_{1},\,\,\,Br_{8}=BR_{6}=0 \\
Br_{5} & = & \alpha_{2},\,\,\,Br_{4}=BR_{2}=0 \nonumber\\
Br_{9} & = & \alpha_{3},\,\,\,Br_{7}=BR_{3}=0 \nonumber
\end{eqnarray}

In the following section, we evaluate the contribution of gravitino decays to the electron-positron spectrum and the EGB by using only this combination between branching fractions and effective parameters.

\section{Statistical analysis setup}
\label{sec:stats}

In order to compare the results about the optimization of the TRpV model with those obtained with the analogous BRpV scenario in our previous work~\cite{Carquin:2015uma}, we start the statistical analysis of TRpV by considering the same sources of data for electron-positron cosmic rays and $\gamma$-rays as in our previous work. This data is composed by the the following sets:

\begin{itemize}
\item[$D_1$:] The positron fraction measured by AMS-02 between 0.5 and 500 GeV~\cite{Accardo:2014lma},
\item[$D_2$:] The independent measurement of the electron and positron fluxes by AMS-02 between 0.5 and 700 GeV~\cite{Aguilar:2014mma},
\item[$D_3$:] The measurement of the sum of electron and positron spectrum measured by AMS-02 between 0.5 GeV and 1 TeV~\cite{Aguilar:2014fea},
\item[$D_4$:] The spectrum of isotropic diffuse gamma-ray emission between 100 MeV and 820 GeV measured by Fermi-LAT~\cite{Ackermann:2014usa}, from which we derive the limits on the EGB,
\end{itemize}

Therefore, in the first step of the analysis, we use the first three data sets together in order to adjust the free parameters of the TRpV scenario. The last one is used to cross check the consistency of the best fit model to the EGB limits. Besides, we also consider other sources of data about electron and positron cosmic rays in order to study the consistency of our previous results when more data is considered. These sets are given by:

\begin{itemize}
\item[$D_5$:] The extended measurement of the sum of electron and positron spectrum by CALET between 11 GeV and 4.8 TeV~\cite{Adriani:2018ktz},
\item[$D_6$:] The direct detection of the spectrum of electrons plus positrons measured by DAMPE between 25 GeV and 4.6 TeV~\cite{Ambrosi:2017wek} and
\end{itemize}
 
Thus, in the second step of the statistical analysis, we adjust the parameters of the TRpV scenario considering different combinations between AMS, CALET and DAMPE data and in every case we cross-check the consistency of the optimized models agaisnt the EGB limits defined from $D_3$. 
 
\subsection*{Likelihood definition and optimization algorithm} 
 
Considering the definition of the standard astrophysical sources of electron and positron signals and the analysis of the gravitino sector of the previous section, we identify a total of eight free parameters that need to be optimized in order to adjust the electron and positron data. These parameters are given by $C_e,\, \gamma_e,\,C_p,\,\gamma_p,\,m_G, \,\tau_G,\,\alpha_1$ and $\alpha_2$. Anyway, we notice that each data set shown above may depend only on a subset of these parameters. 

In order to evaluate the goodness of the fit for a given point in this parameter space, we use a set of Gaussian likelihoods. Since the full likelihood function can be quite long and indeed can vary depending on the data sets considered for the analysis, here we only show the part that corresponds to the positron measurements, which is given by

\begin{equation}
\log  \mathcal{L}_{\text{Positrons}} = -\frac{1}{2} \sum_i{\left( \frac{(\Phi_D^p(E_i) - \Phi_M^p(\theta_p,E_i ))^2}{(\sigma_D^2 + j\times (\Phi_D^{p}(\theta_p,E_i))^2)} - \frac{1}{(\sigma_D^2 + j\times (\Phi_D^{p}(\theta_p,E_i))^2)}  \right) },
\end{equation}

\noindent with $\sigma_D$ the statistical uncertainty of the measurement $D$ and $\Phi_D^p(E_i)$ the observed flux of positrons in the data set $D$. The parameter $j$ increases the nominal uncertainty by a fraction of the model, to account for systematic effects and correlations among different data sets~\cite{Hogg:2010yz}. In this case, the model is defined as

\begin{equation}
\Phi_M^p(\theta_p,E_i ) = \Phi_B^p(C_p, \gamma_P, E_i) + \Phi_{G}^{p}(m_G, \tau_G, \alpha_1, \alpha_2, E_i).
\end{equation}

The likelihood functions considering other measurements can be defined analogously. In order to explore the full likelihood, find the best fit values and credible regions of our unknown parameters, we use the Bayesian inference package Multinest \cite{Feroz:2008xx} through its Python interface \verb|PyMultinest|~\cite{Buchner:2014nha}. Later, we consider the best fit predictions of these scenarios in the $\gamma$-ray region in order to check the consistency of our results with EGB limits.

\section{Results}
\label{sec:results}

In the first step of our statistical analysis we adjust the TRpV parameters by considering the data sets $D_1$, $D_2$ and $D_3$, which is the same data used in our previous work. Indeed, we consider two cases, in the first one we use $D_1+D_2$, such that we consider the AMS-02 measurements of the positron fraction and the independent measurements of electrons and positrons. In the second case, we consider $D_1+D_2+D_3$ in order to discuss the effects of considering the sum of electron plus positron, also measured by AMS-02, as an extra source of information. The summary of our results concerning these two cases are given in Table~\ref{table:best-fit-summary} under the columns Case 1 and Case 2 respectively. 

\begin{figure}[htb]
\begin{center}
\includegraphics[height=14cm,width=14cm,angle=0]{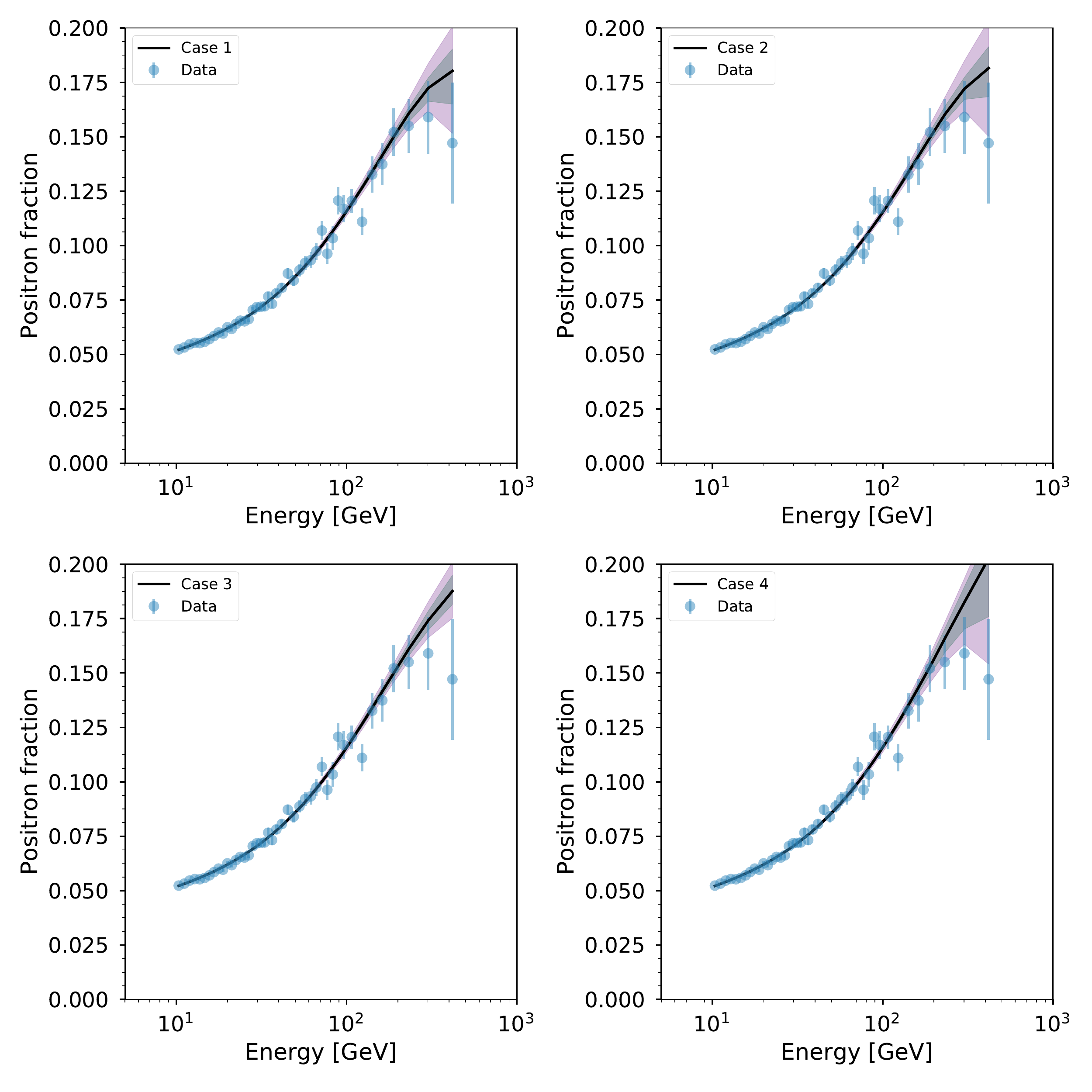}
\caption{Best fit results for positron fraction. The scenarios from Case $1$ to Case $4$ are ordered from left to right and from top to bottom.}
\label{fig:bf-positron-fraction-spectrum}
\end{center}
\end{figure}

\begin{figure}[htb]
\begin{center}
\includegraphics[height=14cm,width=14cm,angle=0]{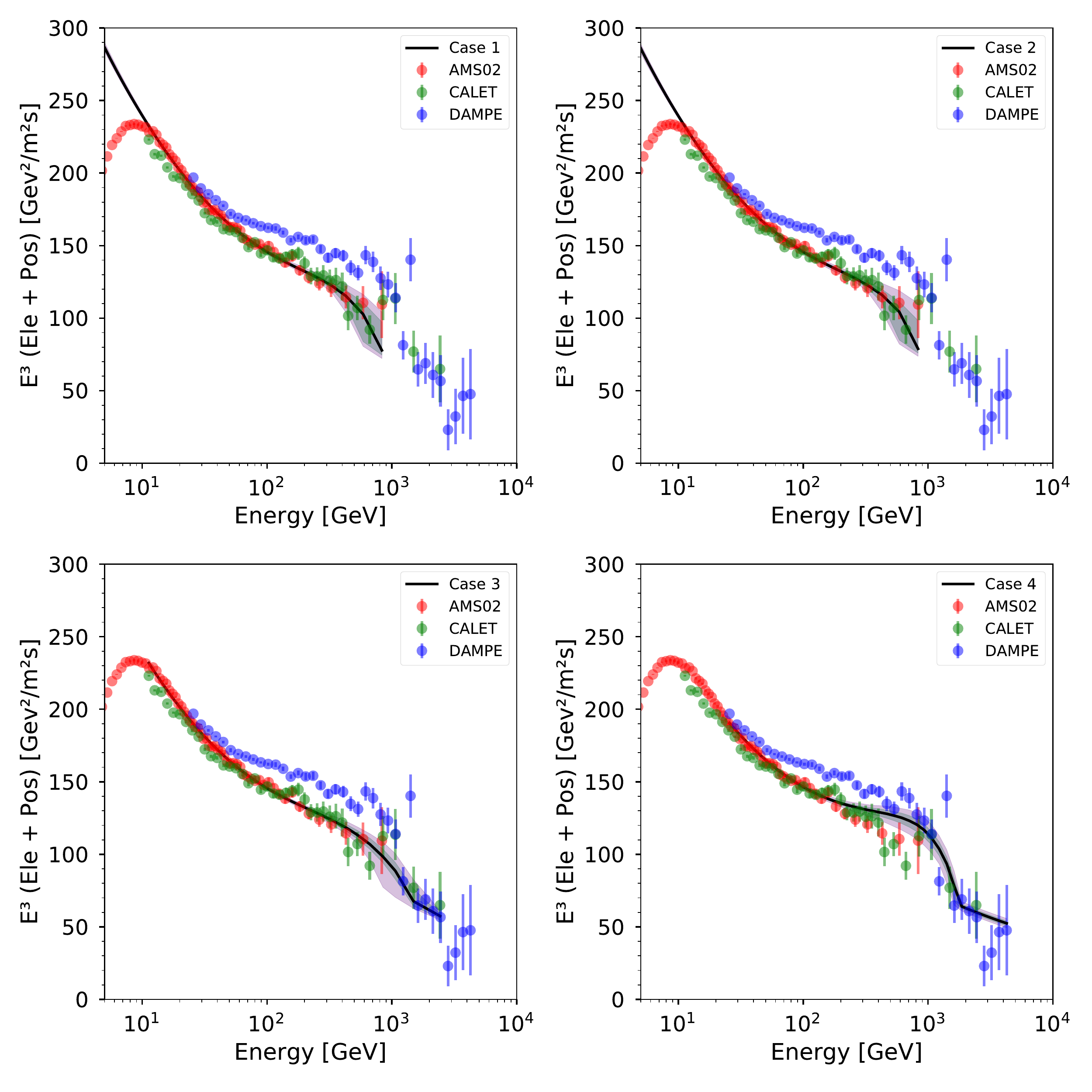}
\caption{Best fit results for electron plus positron flux. The scenarios from Case $1$ to Case $4$ are ordered from left to right and from top to bottom.}
\label{fig:bf-electron-plus-positron-spectrum}
\end{center}
\end{figure}

\begin{figure}[htb]
\begin{center}
\includegraphics[height=12cm,width=14cm,angle=0]{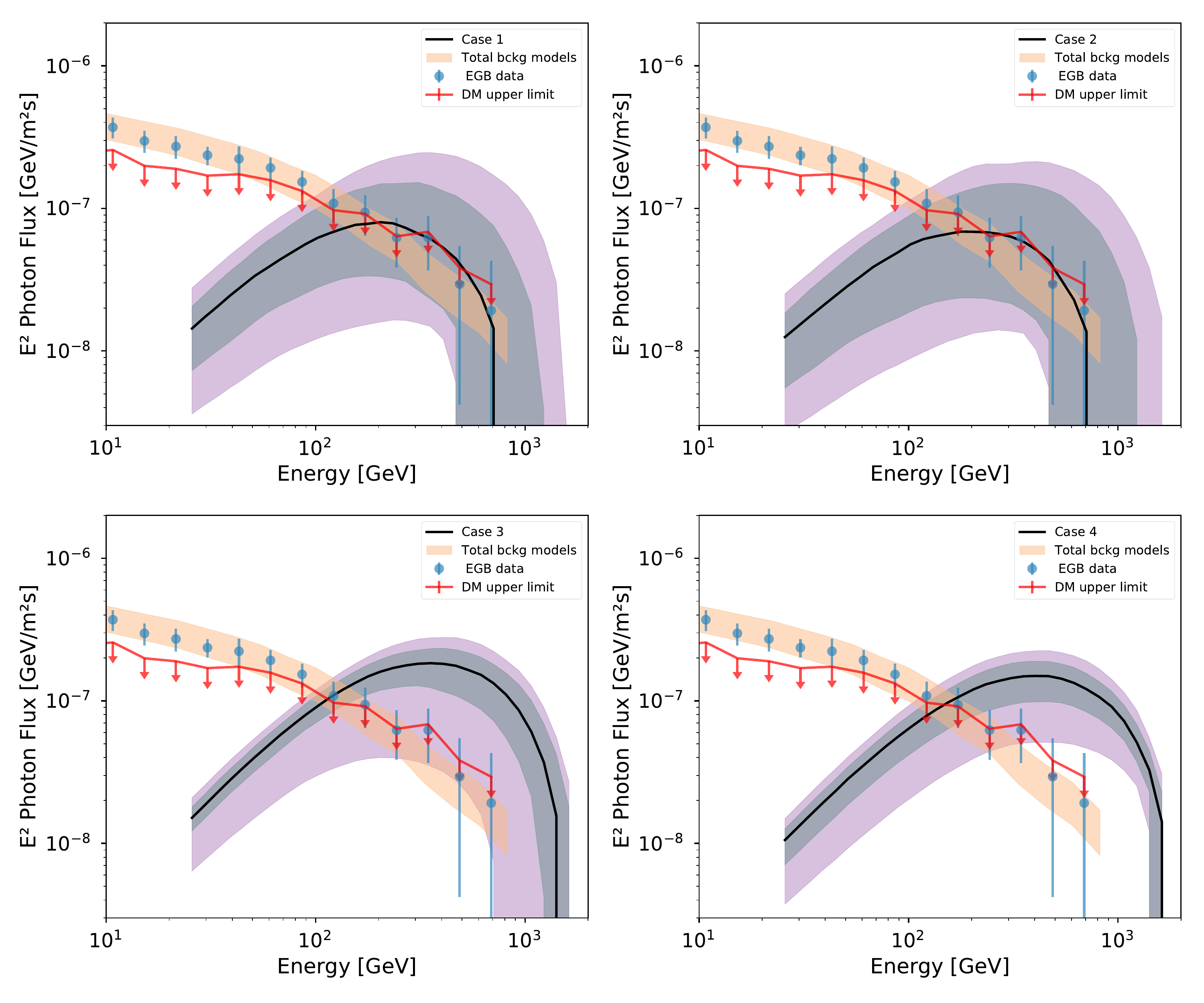}
\caption{Best fit results for photon flux compared to EGB limits.The scenarios from Case $1$ to Case $4$ are ordered from left to right and from top to bottom.}
\label{fig:bf-photon-egb-spectrum}
\end{center}
\end{figure}

From the results of Table~\ref{table:best-fit-summary} we see that in both cases the lifetime of the gravitino is around $4\times10^{26}\,s$ with a strong preference of the channel $\tau^{-}l^{+}\nu$. The best fit lines for Case 1 and 2 for the positron fraction are given in the top panels of Fig.~\ref{fig:bf-positron-fraction-spectrum}. The predictions for the electron plus positron flux in both cases are given in Fig.~\ref{fig:bf-electron-plus-positron-spectrum}. We notice that both cases produce very similar results, which are almost indistinguishable by simple inspection. In Fig.~\ref{fig:bf-photon-egb-spectrum} top panel, we show the comparison between the $\gamma$-ray distributions of both cases and the EGB limits, which are obtained as in our previous work. In these plots we can find our main results, since here we can see that the $\gamma$-ray spectrum obtained in the TRpV scenario is indeed compatible with EGB limits. The central values of the spectrum just touch the upper limit points and almost half the points inside the 95\% C.L. regions are below these limits.

We also notice that the best fit of the gravitino mass can vary from 1.2 to 2.3 TeV when we pass from Case 1 to Case 2. This is probably because the measurement of the electron plus positron signal, included only in Case 2, indeed can reach higher energies than the isolated measurement of electrons and positrons, which in turn may shift the spectrum of  gravitino decays to higher energies. Although this effect is not super clear when we compare the figures for Case 1 and Case 2, we will see that this effect is more notorious when we consider the data about electron plus positron from CALET and DAMPE.

Thus, in the second step of the analysis we adjust the TRpV parameters by considering two new sources of the electron plus positron measurements. The first one (Case 3) considers the combination $D_1+D_2+D_5$ in order to replace the measurement of AMS-02 by the analogous measurements of CALET. In the second case (Case 4) we implement a similar approach by considering DAMPE data. The obtained best fit values of these two cases are given in Table~\ref{table:best-fit-summary} under the columns Case 3 and Case 4. We clearly can see that in both cases the best fit gravitino mass suddenly increases from 2 TeV (Case 2) to almost 4 TeV (Cases 3 and 4). The value of the gravitino lifetime decreases from $4\times10
^{26}\,s$ to $2.3\times10^{26}\,s$ probably in order to balance the change on mass. The preference for the channel $\tau^{-}l^{+}\nu$ is hold it as in previous cases. In Fig.~\ref{fig:bf-positron-fraction-spectrum} we see the corresponding predictions for the positron fraction, which seem compatible with current data. 

In Fig.~\ref{fig:bf-electron-plus-positron-spectrum} we see the predictions for electron plus positron for Case 3 and Case 4, where we can see a big change from the initial two cases, which is explained by the big difference within the observed data itself. More importantly and as consequence of the previous results, we see in Fig.~\ref{fig:bf-photon-egb-spectrum} bottom panels that these two last cases are in conflict with the EGB limits. Since the discrepancy of our predictions mostly differ because of the difference on the considered data, going from allowed to forbidden scenarios, we believe that further conclusions about the TRpV scenario must consider the study about the origin of the discrepancy in the data about electron plus positron measurements. 

\begin{table}[t]
\centering{}%
\begin{tabular}{|c|c|c|c|c|}
\hline 
 Parameter & Case 1 & Case 2 & Case 3 & Case 4 \tabularnewline
\hline 
\hline 
$C_p$ [1/GeV cm$^2$ s str] & 14.90 & 14.74 & 14.93 & 14.37 \tabularnewline
\hline 
$\gamma_p$ & 3.11 & 3.10 & 3.11 & 3.09  \tabularnewline
\hline 
$C_e$ [1/GeV cm$^2$ s str] & 426.10 & 421.77 & 422.08 & 422.67 \tabularnewline
\hline 
$\gamma_e$ & 3.27 & 3.27 & 3.27 & 3.27 \tabularnewline
\hline 
$m_G$ [GeV] & 1281 & 2274 & 3604 & 3751 \tabularnewline
\hline 
$\tau_G$ [10$^{26}$ s] & 4.61 & 3.59 & 2.27 & 2.29  \tabularnewline
\hline 
$\alpha_1: e^-l^+\nu$  & 0.43 &  0.06 & 0.03 & 0.32 \tabularnewline
\hline 
$\alpha_2: \mu^-l^+\nu $  & 0.03 &  0.36 & 0.15 &  0 \tabularnewline
\hline
\hline
$\alpha_3: \tau^-l^+\nu$  & 0.54 &  0.58 & 0.82 & 0.68 \tabularnewline
\hline 
\end{tabular}
\caption{Best fit parameters for the different cases defined in previous section. Recall that $\alpha_3= 1-(\alpha_1+\alpha_2)$ by definition.}
\label{table:best-fit-summary}
\end{table}

\section{Conclusions}
\label{sec:conclusions}

In this work we consider a dark matter model that includes a meta-stable gravitino that can decay to SM particles only though TRpV couplings. This model is introduced in order to adjust the electron-positron cosmic ray data, in particular the anomalous rising in the positron fraction observed by AMS-02 and others. Based on the comparison between decay configurations allowed in TRpV and BRpV, we suggest that in the former it is possible to adjust the electron-positron data in consistency with EGB limits, thus improving the results obtained only with BRpV decays, that overshoot these limits.

From the statistical analysis of the TRpV scenario, considering the data from AMS-02, CALET, DAMPE and Fermi-LAT, we find two important results. On the one hand, we explicitly show that TRpV is indeed able to explain the electron-positron data measured by AMS-02. More importantly, we obtain that these results are compatible with the EGB limits obtained from Fermi-LAT data. This is an improvement on the results obtained by using only BRpV couplings.

On the other hand we show that this compatibility with the EGB limits can be affected by the inclusion of electron-positron data from CALET and DAMPE. Basically, the extended domain of the electron plus positron flux measured by these two experiments shifts the gravitino contribution to higher energies, which is the region constrained by the EGB limits. We notice that the root of this conflict is given by the discrepancy between AMS-02 electron plus positron data and the corresponding data from CALET and DAMPE. Finally, we suggest a deeper study about this discrepancy in order to further analyze the gravitino dark matter with TRpV model and indeed any other dark matter scenario with signals in these regions.  

\section*{Acknowledgments}

{\small 
We acknowledge support from the ANID-Chile grants Basal-CATA
PFB-06/2007 and Basal AFB-170002 (J.B.), FONDECYT Postdoctorados 3160439 (J.B.) and the Ministry of Economy, Development, and Tourism's Millennium Science Initiative through grant IC120009, awarded to The Millennium Institute of Astrophysics, MAS (J.B.). B.P. is currently supported by a Postdoctoral Fellowship from the joint committee ESO-Government of Chile. B.P. also thanks the support of the State of S\~{a}o Paulo Research Foundation (FAPESP) during the initial stages of this work. E.C. is supported by grant; FONDECYT No. 1190886, N.V. is supported by FONDECYT No. 11170109, E.C. and N.V. are also suported by grant ANID PIA/APOYO AFB180002.

}

\appendix
\section{Appendix: discussion about the gravitino lifetime and neutrino masses}
\label{sec:lifetime-neutrinomasses}

The full expressions for the gravitino decay width, considering trilinear
R-Parity violation, are given in~\cite{Moreau:2001sr}. For instance, from
these expressions we can get an approximated formula for the leptonic
decay $\Gamma(\tilde{G}\rightarrow\nu_{i}e_{j}\bar{e}_{k})$ by assuming
that the mass of the sleptons mediating the three body decay are
equal, such that $m_{\tilde{\nu}_{iL}}=m_{\tilde{e}_{jL}}=m_{\tilde{e}_{kR}}=\tilde{m}$,
and expand in taylor series around the variable $m_{G}/\tilde{m}$
to obtain

\begin{equation}
\Gamma(\tilde{G}\rightarrow\nu_{i}e_{j}\bar{e}_{k})\approx\frac{1}{96(2\pi)^{3}}\frac{\lambda_{ijk}^{2}}{8M_{\star}^{2}}\frac{m_{G}^{7}}{\tilde{m}^{4}},\label{eq:GravDecayApp}
\end{equation}

\begin{figure}
\begin{centering}
\includegraphics[scale=1.2]{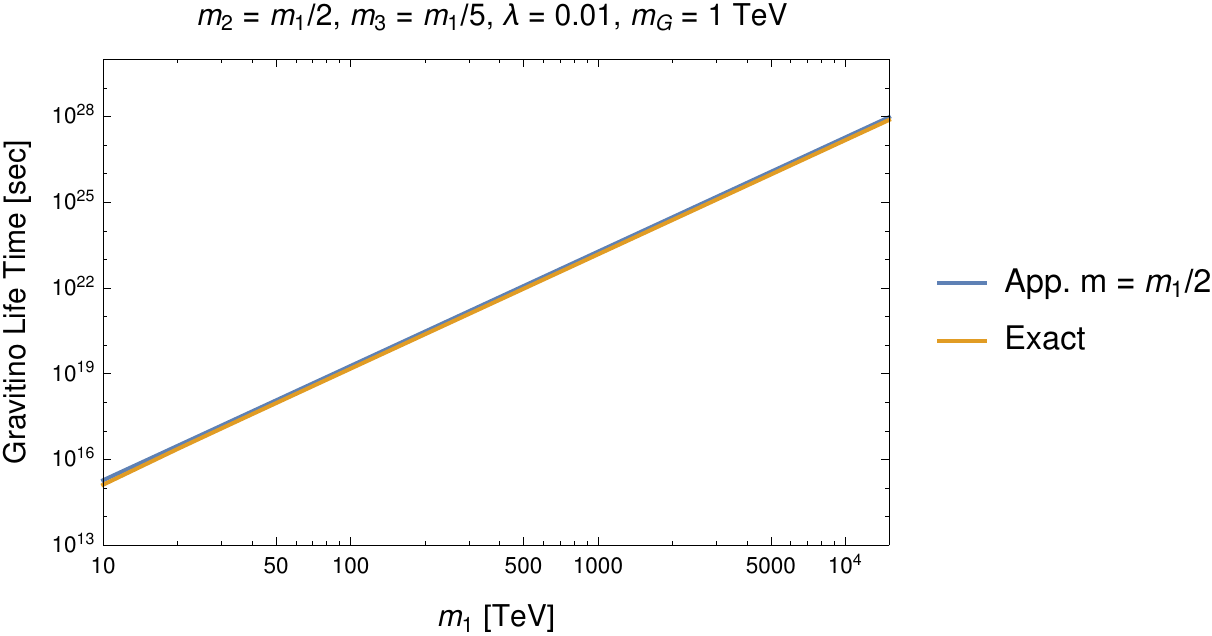}
\par\end{centering}

\caption{\label{fig:Gravitino-life-time}Gravitino life time in Trilinear RpV
for $\lambda_{ijk}=0.01,\,m_{G}=1\,\mbox{TeV}$. For simplicity we
use $m_{1},\,m_{2},\,m_{3}$ and $m$ instead of $m_{\tilde{\nu}_{iL}},\,m_{\tilde{e}_{jL}},\,m_{\tilde{e}_{kR}}$
and $\tilde{m}$.}
\end{figure}

This result shows that the decay width
(lifetime) decreases (increases) rapidly as we increase $\tilde{m}$,
as expected. We expect that a similar behavior should be obtained
even when the mass of sleptons are not equal. Indeed, we have verified this expectation numerically, by evaluating
the full expression given in~\cite{Moreau:2001sr} using the maximum numerical
precision in Mathematica. For instance, in Fig. \ref{fig:Gravitino-life-time}
we plot the gravitino lifetime as a function of $m_{\tilde{\nu}_{iL}}$
for $m_{\tilde{e}_{jL}}=m_{\tilde{\nu}_{iL}}/2$ and $m_{\tilde{e}_{kR}}=m_{\tilde{\nu}_{iL}}/5$.
Also, in the same figure we plot the lifetime derived from Eq. \ref{eq:GravDecayApp}
evaluated at $\tilde{m}=m_{\tilde{\nu}_{iL}}/2$ in order to check
that both approaches, exact computation and approximated formula,
behave similarly. Therefore, the gravitino lifetime can be written as in Eq.~\ref{eq:GravLifeTime},

\begin{equation}
\tau_{G} \approx 10^{26}\,\text{s}\,\left(\frac{1}{\lambda_{ijk}\lambda_{ijk}}\right)\left(\frac{\tilde{m}}{2\times 10^{7}\mbox{\,GeV}}\right)^{4}\left(\frac{1\,\mbox{TeV}}{m_{G}}\right)^{7}\label{eq:GravLifeTime}
\end{equation}

\noindent where we have normalized with respect to $10^{26}$ s
since this is the order of magnitude required by experiments such
as AMS-02 and Fermi-LAT in order to fit the electron positron data
in the first case or to avoid gamma ray constraints in the second.


In TRpV the neutrino mass matrix receives contributions from
1-loop diagrams that contain both a charged lepton and the corresponding
slepton. Indeed, we have derived the following expression

\begin{eqnarray*}
M_{ij}^{\nu\,(1)} & \approx & \frac{1}{16\pi^{2}}\sum_{gr}s_{\tilde{l}}c_{\tilde{l}}(\lambda_{igr}\lambda_{jrg}+\lambda_{jgr}\lambda_{irg})m_{g}\ln\frac{m_{\tilde{l}_{r2}}^{2}}{m_{\tilde{l}_{r1}}^{2}}
\end{eqnarray*}

\noindent where $i$ and $j$ are neutrino generation indices that
runs from 1 to 3. $g$ is a charged lepton index that also run from
1 to 3, as well as $r$ which is a slepton index. Thus, it can be
seen that for order one parameters, $s_{\tilde{l}} \sim c_{\tilde{l}} \sim \ln(m_{l_{r2}}^{2}/m_{l_{r1}}^{2})\sim 1$, we can get neutrino masses around the eV scale for $\lambda_{ijk}\approx0.01$
even for $m_{g}\approx m_{e}$. Indeed, by following the expressions given in \cite{Chun:2004mu} for the
contribution of $\lambda'$ trilinear terms, we can get by analogy
that the dominant term in the leptonic sector is 

\begin{eqnarray*}
M_{ij}^{\nu\,(1)} & \approx & \frac{1}{8\pi^{2}}\lambda_{i23}\lambda_{j32}\frac{m_{\mu}m_{\tau}A_{\tau}}{\tilde{m}^{2}}\\
 & \approx & 2\times10^{-2}\mbox{eV}\,\lambda_{i23}\lambda_{j32}\,\left(\frac{10^{8}\mbox{\,GeV}}{\tilde{m}}\right)\\
 & \approx & 2 \times10^{-2}\mbox{eV}\,(\lambda_{i23}\lambda_{j32})^{1/4}\left(\frac{\tau_{G}}{10^{\text{26}}\,\text{s}}\right)^{1/4}\left(\frac{m_{G}}{2\,\mbox{TeV}}\right)^{7/4}
\end{eqnarray*}

\noindent where $A_{\tau}$ is a free parameter that can be considered
of order $\tilde{m}$, as it is done in~\cite{Chun:2004mu}. Thus, if we
consider this formula together with Eq.~(\ref{eq:GravLifeTime}) we
see that we can have contributions to the neutrino mass matrix of
order $10^{-2}$ eV for trilinear couplings of order one, scalar mass scale around $10^8$ GeV and $\tau_{G}\approx10^{26}$ s.

\section{Complementary results from the fit}

In this section we show the complementary plots about electron and positron spectra concerning the statistical analysis of Cases 1 to 4, which are detailed in the main text. 

\begin{figure}[!ht]
\begin{center}
\includegraphics[height=14cm,width=14cm,angle=0]{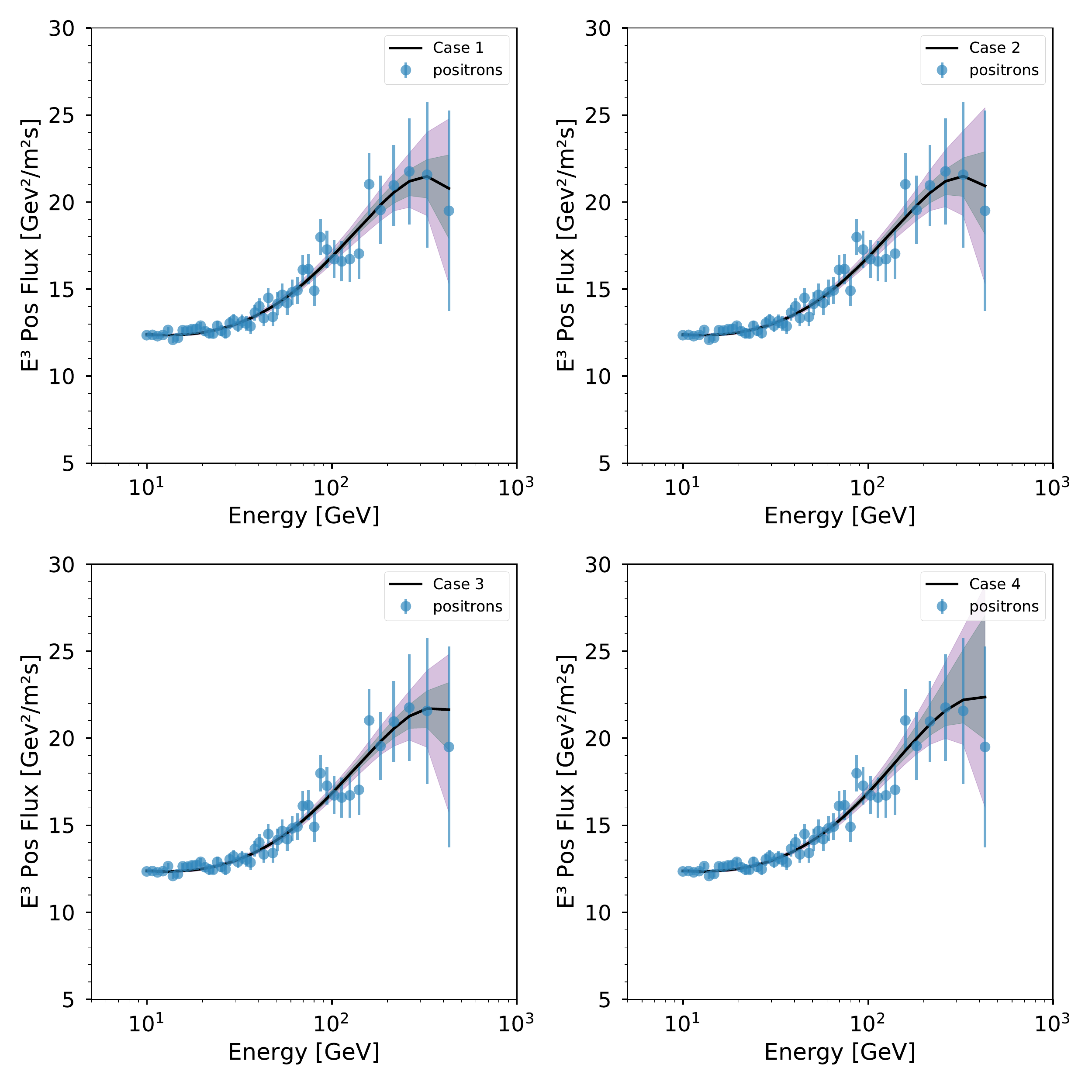}
\caption{Best fit results for positron flux. The scenarios from Case 1 to Case 4 are ordered from left to right and from top to bottom.}
\label{fig:bf-positron-spectrum}
\end{center}
\end{figure}

\begin{figure}[htb]
\begin{center}
\includegraphics[height=14cm,width=14cm,angle=0]{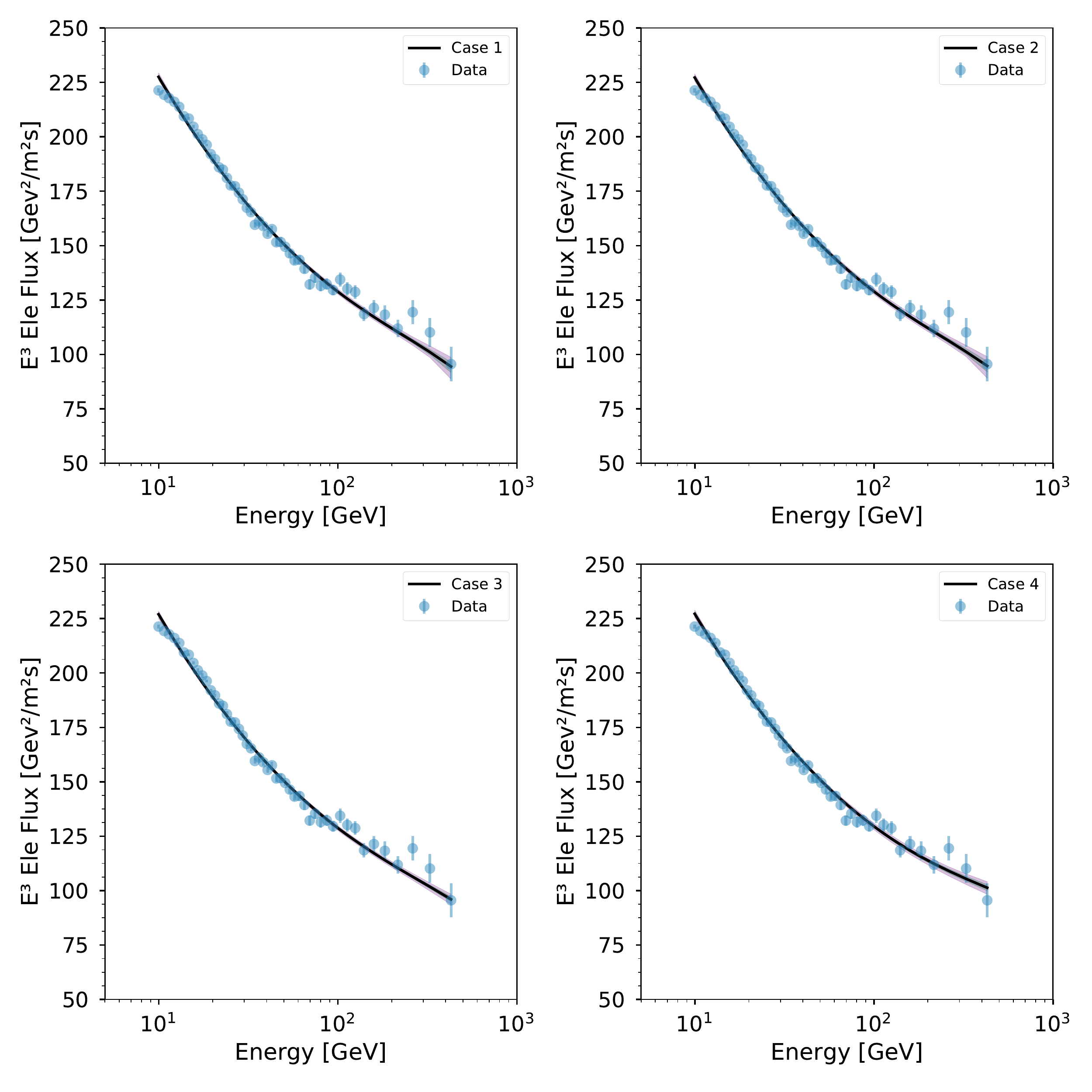}
\caption{Best fit results for electron flux. The scenarios from Case 1 to Case 4 are ordered from left to right and from top to bottom.}
\label{fig:bf-electron-spectrum}
\end{center}
\end{figure}

\clearpage
\bibliographystyle{JHEP}
\bibliography{Gravitino_Trilinear}

\end{document}

%% file: authors.tex
\author[a,b]{Johannes~Buchner,}
\author[d]{Edson~Carquin,}
\author[c]{Marco~A.~D\'\i az,}
\author[a]{Germ\'an~A.~G\'omez-Vargas,\footnote{Now Corporate Data Scientist at \href{https://www.derco.cl/}{Derco},}}
\author[a]{Boris~Panes,}
\author[d]{Nicol\'as Viaux}

\affiliation[a]{Instituto de Astrof\'isica, Pontificia Universidad Cat\'olica de Chile, Avenida Vicu\~na Mackenna 4860, Santiago, Chile}
\affiliation[b]{Millenium Institute of Astrophysics, Vicu\~na MacKenna 4860, 7820436 Macul, Santiago, Chile}
\affiliation[c]{Instituto de F\'isica, Pontificia Universidad Cat\'olica de Chile, Avenida Vicu\~na Mackenna 4860, Santiago, Chile}
\affiliation[d]{Departamento de F\'isica y Centro Cient\'ifico- Tecnol\'ogico de Valpara\'iso (CCTVal), Universidad T\'ecnica Federico Santa Mar\'ia, Av. Espa\~na 1680, Valpara\'iso, Chile}

\emailAdd{johannes.buchner.acad@gmx.com} 
\emailAdd{edson.carquin@usm.cl}
\emailAdd{mad@susy.fis.puc.cl}
\emailAdd{ggomezv@uc.cl}
\emailAdd{bpanes@astro.puc.cl}
\emailAdd{nicolas.viaux@usm.cl}

%% file: Gravitino_Trilinear.bbl
\providecommand{\href}[2]{#2}\begingroup\raggedright\begin{thebibliography}{10}

\bibitem{Adriani:2008zr}
{\bf PAMELA} Collaboration, O.~Adriani et~al., {\it {An anomalous positron
  abundance in cosmic rays with energies 1.5-100 GeV}},  {\em Nature} {\bf 458}
  (2009) 607--609, [\href{http://xxx.lanl.gov/abs/0810.4995}{{\tt
  arXiv:0810.4995}}].

\bibitem{Accardo:2014lma}
{\bf AMS} Collaboration, L.~Accardo et~al., {\it {High Statistics Measurement
  of the Positron Fraction in Primary Cosmic Rays of 0.5500 GeV with the Alpha
  Magnetic Spectrometer on the International Space Station}},  {\em Phys. Rev.
  Lett.} {\bf 113} (2014) 121101.

\bibitem{Aguilar:2014mma}
{\bf AMS} Collaboration, M.~Aguilar et~al., {\it {Electron and Positron Fluxes
  in Primary Cosmic Rays Measured with the Alpha Magnetic Spectrometer on the
  International Space Station}},  {\em Phys. Rev. Lett.} {\bf 113} (2014)
  121102.

\bibitem{Aguilar:2014fea}
{\bf AMS} Collaboration, M.~Aguilar et~al., {\it {Precision Measurement of the
  ($e^+ + e^−$) Flux in Primary Cosmic Rays from 0.5 GeV to 1 TeV with the
  Alpha Magnetic Spectrometer on the International Space Station}},  {\em Phys.
  Rev. Lett.} {\bf 113} (2014) 221102.

\bibitem{Adriani:2018ktz}
O.~Adriani et~al., {\it {Extended Measurement of the Cosmic-Ray Electron and
  Positron Spectrum from 11 GeV to 4.8 TeV with the Calorimetric Electron
  Telescope on the International Space Station}},  {\em Phys. Rev. Lett.} {\bf
  120} (2018), no.~26 261102, [\href{http://xxx.lanl.gov/abs/1806.0972}{{\tt
  arXiv:1806.0972}}].

\bibitem{Ambrosi:2017wek}
{\bf DAMPE} Collaboration, G.~Ambrosi et~al., {\it {Direct detection of a break
  in the teraelectronvolt cosmic-ray spectrum of electrons and positrons}},
  {\em Nature} {\bf 552} (2017) 63--66,
  [\href{http://xxx.lanl.gov/abs/1711.1098}{{\tt arXiv:1711.1098}}].

\bibitem{Ackermann:2014usa}
{\bf Fermi-LAT} Collaboration, M.~Ackermann et~al., {\it {The spectrum of
  isotropic diffuse gamma-ray emission between 100 MeV and 820 GeV}},  {\em
  Astrophys. J.} {\bf 799} (2015) 86,
  [\href{http://xxx.lanl.gov/abs/1410.3696}{{\tt arXiv:1410.3696}}].

\bibitem{Hooper:2018kfv}
D.~Hooper, {\it {TASI Lectures on Indirect Searches For Dark Matter}},
  \href{http://xxx.lanl.gov/abs/1812.0202}{{\tt arXiv:1812.0202}}.

\bibitem{Grefe:2008zz}
M.~Grefe, {\it {Neutrino signals from gravitino dark matter with broken
  R-parity}},  \href{http://xxx.lanl.gov/abs/1111.6041}{{\tt arXiv:1111.6041}}.

\bibitem{2012PhRvD..86h3506C}
M.~{Cirelli}, E.~{Moulin}, P.~{Panci}, P.~D. {Serpico}, and A.~{Viana}, {\it
  {Gamma ray constraints on decaying dark matter}},  {\em \prd} {\bf 86} (Oct.,
  2012) 083506, [\href{http://xxx.lanl.gov/abs/1205.5283}{{\tt
  arXiv:1205.5283}}].

\bibitem{Ando:2015qda}
S.~Ando and K.~Ishiwata, {\it {Constraints on decaying dark matter from the
  extragalactic gamma-ray background}},  {\em JCAP} {\bf 1505} (2015), no.~05
  024, [\href{http://xxx.lanl.gov/abs/1502.0200}{{\tt arXiv:1502.0200}}].

\bibitem{Laletin:2016egv}
M.~Laletin, {\it {A no-go theorem for the dark matter interpretation of the
  positron anomaly}},  {\em Frascati Phys. Ser.} {\bf 63} (2016) 7--12,
  [\href{http://xxx.lanl.gov/abs/1607.0204}{{\tt arXiv:1607.0204}}].

\bibitem{Liu:2016ngs}
W.~Liu, X.-J. Bi, S.-J. Lin, and P.-F. Yin, {\it {Constraints on dark matter
  annihilation and decay from the isotropic gamma-ray background}},  {\em Chin.
  Phys.} {\bf C41} (2017), no.~4 045104,
  [\href{http://xxx.lanl.gov/abs/1602.0101}{{\tt arXiv:1602.0101}}].

\bibitem{Belotsky:2016tja}
K.~Belotsky, R.~Budaev, A.~Kirillov, and M.~Laletin, {\it {Fermi-LAT kills dark
  matter interpretations of AMS-02 data. Or not?}},  {\em JCAP} {\bf 1701}
  (2017), no.~01 021, [\href{http://xxx.lanl.gov/abs/1606.0127}{{\tt
  arXiv:1606.0127}}].

\bibitem{Carquin:2015uma}
E.~Carquin, M.~A. Diaz, G.~A. Gomez-Vargas, B.~Panes, and N.~Viaux, {\it
  {Confronting recent AMS-02 positron fraction and Fermi-LAT extragalactic
  $\gamma$-ray background measurements with gravitino dark matter}},  {\em
  Phys. Dark Univ.} {\bf 11} (2016) 1--10,
  [\href{http://xxx.lanl.gov/abs/1501.0593}{{\tt arXiv:1501.0593}}].

\bibitem{Grefe:2011dp}
M.~Grefe, {\it {Unstable Gravitino Dark Matter - Prospects for Indirect and
  Direct Detection}},  \href{http://xxx.lanl.gov/abs/1111.6779}{{\tt
  arXiv:1111.6779}}.

\bibitem{Moreau:2001sr}
G.~Moreau and M.~Chemtob, {\it {R-parity violation and the cosmological
  gravitino problem}},  {\em Phys.Rev.} {\bf D65} (2002) 024033,
  [\href{http://xxx.lanl.gov/abs/hep-ph/0107286}{{\tt hep-ph/0107286}}].

\bibitem{Chun:2004mu}
E.~J. Chun and S.~C. Park, {\it {Neutrino mass from R-parity violation in split
  supersymmetry}},  {\em JHEP} {\bf 01} (2005) 009,
  [\href{http://xxx.lanl.gov/abs/hep-ph/0410242}{{\tt hep-ph/0410242}}].

\bibitem{Giudice:2004tc}
G.~Giudice and A.~Romanino, {\it {Split supersymmetry}},  {\em Nucl. Phys. B}
  {\bf 699} (2004) 65--89, [\href{http://xxx.lanl.gov/abs/hep-ph/0406088}{{\tt
  hep-ph/0406088}}]. [Erratum: Nucl.Phys.B 706, 487--487 (2005)].

\bibitem{Aguilar:2013qda}
{\bf AMS Collaboration} Collaboration, M.~Aguilar et~al., {\it {First Result
  from the Alpha Magnetic Spectrometer on the International Space Station:
  Precision Measurement of the Positron Fraction in Primary Cosmic Rays of
  0.5-350 GeV}},  {\em Phys.Rev.Lett.} {\bf 110} (2013) 141102.

\bibitem{Hogg:2010yz}
D.~W. Hogg, J.~Bovy, and D.~Lang, {\it {Data analysis recipes: Fitting a model
  to data}},  \href{http://xxx.lanl.gov/abs/1008.4686}{{\tt arXiv:1008.4686}}.

\bibitem{Feroz:2008xx}
F.~Feroz, M.~Hobson, and M.~Bridges, {\it Multinest: an efficient and robust
  bayesian inference tool for cosmology and particle physics},  {\em Mon. Not.
  Roy. Astron. Soc.} {\bf 398} (2009) 1601--1614,
  [\href{http://xxx.lanl.gov/abs/0809.3437}{{\tt arXiv:0809.3437}}].

\bibitem{Buchner:2014nha}
J.~Buchner, A.~Georgakakis, K.~Nandra, L.~Hsu, C.~Rangel, M.~Brightman,
  A.~Merloni, M.~Salvato, J.~Donley, and D.~Kocevski, {\it {X-ray spectral
  modelling of the AGN obscuring region in the CDFS: Bayesian model selection
  and catalogue}},  {\em Astron. Astrophys.} {\bf 564} (2014) A125,
  [\href{http://xxx.lanl.gov/abs/1402.0004}{{\tt arXiv:1402.0004}}].

\end{thebibliography}\endgroup
